\newcommand {\source} {MACS\,J0140.0--0555}

\documentclass[usenatbib,useAMS]{mn2e}

\usepackage{graphicx,times}
\usepackage{longtable,deluxetable}

\title[An X-ray/optical study of MACS\,J0140.0--0555]
  {An X-ray/optical study of the geometry and dynamics of MACS\,J0140.0--0555, a massive post-collision cluster merger \thanks{Some of the data presented herein were obtained at the W.M. Keck Observatory, which is operated as a scientific partnership among the California Institute of Technology, the University of California and the National Aeronautics and Space Administration. The Observatory was made possible by the generous financial support of the W.M. Keck Foundation.} }

\author[Ho, Ebeling \& Richard]
  {I-Ting Ho$^1$, Harald Ebeling$^1$ and Johan Richard$^2$\\
  $^1$Institute for Astronomy, University of Hawaii, 2680 Woodlawn Drive, Honolulu, HI 96822, USA \\
  $^2$Centre de Recherche Astrophysique de Lyon, Universit\'e Lyon 1, 9 Avenue Charles Andr\'e, 69561 Saint Genis Laval Cedex, France}

\begin{document}

\label{firstpage}

\maketitle

\begin{abstract}

We investigate the physical properties, geometry and  dynamics of the massive cluster merger \source\ ($z{=}0.451$) using X-ray and optical diagnostics.  Featuring two galaxy overdensities separated by about 250~kpc in projection on the sky, and a single peak in the X-ray surface brightness distribution located between them, \source\ shows the tell-tale X-ray/optical morphology of a binary, post-collision merger.  Our spectral analysis of the X-ray emission, as measured by our {\it Chandra} ACIS-I observation of the system, finds the intra-cluster medium to be close to isothermal ($\sim$8.5~keV) with no clear signs of cool cores or shock fronts. Spectroscopic follow-up of galaxies in the field of \source\ yields a velocity dispersion of $875^{+70}_{-100}\rm~km~s^{-1}$ ($n_{\rm z}=66$) and no significant evidence of bimodality or substructure along the line of sight. In addition, the difference in radial velocity between the brightest cluster galaxies of the two sub-clusters of $144\pm25\rm~km~s^{-1}$ is small compared to typical collision velocities of several 1000 km s$^{-1}$. A strongly lensed background galaxy at $z{=}0.873$ (which features variable X-ray emission from an active nucleus) provides the main constraint on the mass distribution of the system. We measure $M({<}75 {\rm kpc}) = (5.6\pm 0.5)\times 10^{13}$ M$_\odot$ for the north-western cluster component and a much less certain estimate of $(1.5-3)\times 10^{13}$ M$_\odot$ for the south-eastern subcluster. These values are in good agreement with our X-ray mass estimates which yield a total mass of \source\ of $M({<}r_{\rm 500}) \sim (6.8-9.1)\times 10^{14}$ M$_\odot$.

Although all optical and X-ray properties of \source\ are consistent with a well advanced head-on merger proceeding along an axis close to the plane of the sky, the degeneracy between Hubble flow and peculiar velocity prevents us from obtaining a quantitative constraint on the inclination angle of the merger axis. The lack of pronounced substructure in the cluster gas distribution and the proximity of the two optical cluster cores both in projection and in radial velocity suggests that the merger is observed well after the primary collision and, possibly, after turnaround. A weak-lensing analysis of the mass distribution in \source\ has the potential of yielding constraints on the self-interaction cross section of dark matter similar to those obtained for MACSJ\,0025.4$-$1222, although the smaller projected separation of the two cluster components makes the measurement more challenging.

\end{abstract}


\section{Introduction}

As the largest gravitationally bound systems in the universe, clusters of galaxies harbour copious amounts of both luminous and dark matter.  Located at the densest regions of the Cosmic Web \citep{bon96}, clusters grow through semi-continuous accretion of matter along large-scale filaments as well as through discrete mergers with nearby clusters. The dynamics of both of these processes are well explored by numerical simulations \citep[e.g.,][]{roe96,col99,col05,rit02,cec11}. Detailed studies of cluster mergers thus provide a unique opportunity for studies of the physics of mass assembly at the nodes of the Cosmic Web, as well as of the dynamical behaviour of dark matter. 

A crucial and obvious prerequisite for such studies is the identification of massive cluster mergers, either active or very recent. A shortlist of extreme, but geometrically simple, linear mergers was recently compiled by \citet{man12} by examining the X-ray and optical morphology of a statistical sample of very X-ray luminous clusters at $z>0.15$. By virtue of being binary, head-on mergers, these systems should be particularly well suited for quantitative interpretation of the immediate observables. The latter include a simple, yet powerful diagnostic that can be used to investigate the nature and properties of dark matter, namely the spatial separation of collisional and non-collisional cluster components during cluster mergers. If dark matter is indeed collisionless \citep[e.g., ][]{gne01,nat02}, cluster mergers should result in a clear segregation between the collisionless dark matter and the viscous intra-cluster medium (ICM) which is subject to shock heating and compression.

Extreme cases of a segregation of this kind were detected in 1E0657--56 \citep[$z{=}0.296$; the Bullet Cluster; ][]{clo04,clo06}, and MACS\,J0025.4--1222 \citep[$z{=}0.586$; ][]{bra08}. Although alternative explanations are still being explored \citep[e.g., modified Newtonian dynamics and tensor-vector-scalar gravity; ][]{ang07,fei08}, the observed offsets between the centroid of the X-ray emission and the distribution of the total gravitational mass as reconstructed from weak-lensing observations provides strong and direct empirical evidence of the existence of dark matter, and allows upper limits to be placed on the dark-matter self-interaction cross-section, $\sigma/m$ \citep{mar04,ran08}. Of the various methods available to place observational limits on $\sigma/m$ from cluster mergers, direct measurements of the mass surface density distribution in a post-collision, head-on, equal-mass merger are perhaps the most powerful as they can be obtained straightforwardly for any massive merger, as long as the merging components are well separated on the sky \citep{mar04,bra08}. 

We here use X-ray and optical observations of \source, one system from the list of \citet{man12}, in order to unravel the dynamics and geometry of the merger, and to establish its usefulness for measurements of $\sigma/m$.

\subsection{\source}
\begin{figure}
\includegraphics[width=80mm]{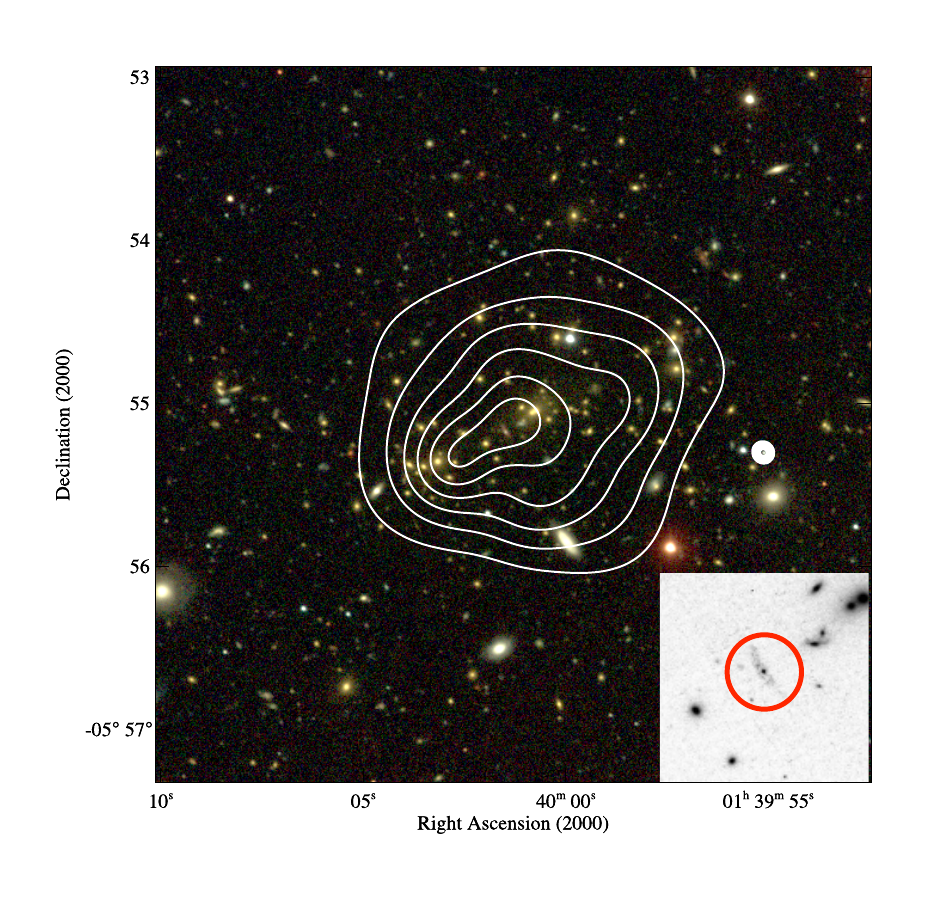}
\caption{Iso-intensity contours of the adaptively smoothed \citep{ebe06} X-ray emission from \source\ as observed with {\it Chandra} (ACIS-I; 10~ks exposure) overlaid on an optical image (V, R, I) obtained with the UH2.2m telescope. Contour levels are spaced logarithmically. The inset shows a putative gravitational arc near the peak of the X-ray emission, detected by an {\it HST} snapshot image taken for program GO-10491 (PI Ebeling).}\label{intro_source}
\end{figure}

\source\ ($z{=}0.451$)  is one of several clusters from the MACS sample \citep{ebe01,ebe07,ebe10} with an X-ray/optical morphology similar to that of MACS\,J0025.4--1222.  In the optical, \source\ clearly comprises two sub-clusters separated  by about 250\,kpc in projection on the sky. At X-ray wavelengths, a short (10\,ks) {\it Chandra} observation revealed extended  emission from the gaseous ICM elongated along the axis connecting the brightest cluster galaxies (BCGs) of the two sub-clusters and featuring a single peak {\it between} the BCGs (Fig.~\ref{intro_source}). Consequently, the system was selected by \citet{man12} as one of eleven likely post-collision, binary, head-on mergers. A snapshot image of the system (see inset in Fig.~\ref{intro_source}), taken with the Advanced Camera for Surveys (ACS) aboard the Hubble Space Telescope ({\it HST})  shows a putative gravitational arc close to the peak of the extended ICM emission. If confirmed, this arc would yield immediate constraints on the mass of one, and possibly of both of the two apparent subclusters.

In order to confirm this tentative picture and to obtain a quantitative description of the merger history and physical properties of \source, we conducted additional observations with {\it Chandra} and {\it HST}, as well as groundbased galaxy spectroscopy with Keck telescopes. 
In this paper, we present the results of our comprehensive analysis of all data collected. Specifically, we perform spatial and spectral analyses of the {\it Chandra} X-ray data to quantify the distribution, mass, and temperature of the ICM. In addition, we assess the distribution and dynamics of the cluster galaxy population based on groundbased spectroscopy. The latter is also used to obtain redshifts of putative gravitational arcs that may yield strong-lensing constraints on the dark matter distribution in \source. The paper is structured as follows. Section 2 describes the observations and data reduction. Section 3 presents the results from of our X-ray and optical analyses. In Section 4, we discuss these results and their implications for the nature of the merger. A summary is given in Section 5.

Throughout this paper, we assume the concordance $\Lambda$CDM cosmology with $\rm H_0=70~km~s^{-1}~Mpc^{-1}$, $\rm\Omega_M=0.3$, and $\Omega_\Lambda=0.7$, for which 1\arcsec\ corresponds to 5.77\,kpc at the cluster redshift of $z{=}0.451$. In addition, abundances are measured using the abundance table supplied by \citet{and89}. All errors are $1\sigma$ unless stated otherwise.

\section{observation and data reduction}
\subsection{X-ray data}
\source\ was observed with the Advanced CCD Imaging Spectrometer  \citep[ACIS,][]{gar03} aboard the {\it Chandra} X-ray observatory in 2004 June (ObsID 5013) and 2010 September (ObsID 12243) for a total exposure time of 10.3~ks and 19.6~ks respectively in VFaint mode. The data were reduced following standard {\it Chandra} data reduction procedures\footnote{http://cxc.harvard.edu/ciao/index.html} using {\it CIAO} 4.3 and {\it CALDB} 4.4.2 \citep{fru06}. No filtering was necessary as no obvious flares occurred during either observation. For the spatial analysis described in detail below, we merged the two event files after proper alignment. The spectral fitting, however, was performed separately (but simultaneously) on the two event files.

In order to avoid instrumental artifacts caused by position-dependent charge transfer inefficiency, we did not measure the background from an off-source region of our actual datasets, but instead used ``blank-sky" backgrounds drawn from the calibration database. We ensured that the backgrounds from the ``blank-sky'' are at the same levels as the observations by comparing photon count rates at high energy (10.1--11 \,keV) where emission predominately comes from background. The background datasets for the two observations were treated as the science datasets, i.e. they were merged for the spatial analysis, but dealt with separately during the spectral fitting. 

In all fits to the data, spatial or spectral, point sources were excluded.

\subsection{Optical data}

\subsubsection{Hubble Space Telescope imaging}

\source\ was observed in the F606W filter with the Advanced Camera for Surveys \citep[ACS, ][]{gia04} aboard the Hubble Space Telescope on October 3, 2005 for 1,200 seconds as part of Snapshot programme GO-10491 (PI: Ebeling). The data were corrected for CTI effects using the pixel-based algorithm developed by \citet{and10} for this purpose, then multi-drizzled \citep{koe02} and resampled \citep{rho07} into 0.03\arcsec pixels. A second observation was performed on September 12, 2011 with the Wide Field Camera 3 (WFC3) in the F110W and F140W filters both for 706 seconds as part of Snapshot programme GO-12166 (PI Ebeling).

\subsubsection{Groundbased spectroscopy}

Follow-up spectroscopy of likely cluster galaxies and potential strong-lensing features in \source\ were performed with the Low Resolution Imaging Spectrograph \citep[LRIS, ][]{oke95,mcc98,roc10} on the 10m Keck-I telescope on Mauna Kea on December 10, 2010. Using the 600 line/mm grating blazed at 7500\AA\ on the red arm of the spectrograph, the 6800\AA\ dichroic as a beam splitter, and the 300/5000 grism on the blue arm, a multi-object spectroscopy mask with 38 science slits was observed for $3\times 1900$ seconds. Standard data reduction procedures (bias subtraction, flat fielding, sky subtraction, wavelength calibration) were applied before the extraction of one-dimensional spectra for each target object. Twenty eight objects yielded useful spectra for further analyses. Additional spectroscopic follow-up observations were performed on September 11, 2011 with the DEep Imaging Multi-Object Spectrograph \citep[DEIMOS;][]{fab03} on Keck-II telescope, where we used the GG455 order-blocking filter and the Zerodur 600 line/mm grating centred at 6300\AA. Four slit masks containing a total of 60 science slits were designed for this observation and observed for $3\times 600$s; 44 of them yielded useful spectra after standard data reduction procedures were applied. Both data reductions are performed using our pipeline adapted from the Deep Extragalactic Evolutionary Probe 2 Survey data reduction pipeline \citep{dav03}.

\section{Data analysis and results}
\subsection{X-ray surface brightness}\label{sec:xsb}

We quantify the X-ray surface brightness distribution (see Fig.~\ref{intro_source}) by fitting the data with the ``isothermal King" model \citep[also known as $\beta$-model;][]{cav76}.  In the simple spherical case, the gas density profile $\rho(r)$ and the {\it projected} surface brightness profile $S(\hat{r})$ can then be described by 
$$
\rho(r) =\rho_0 [1+({r \over r_0})^2]^{-{3\over2}\beta} 
$$and
$$
S(\hat{r})=S_0 [1+({\hat{r} \over r_0})^2]^{-3\beta+0.5}, 
$$
where $\hat{r}$ is radius $r$ in projection on the sky, $\rho_0$ is the central gas density, $S_0$ is the central surface brightness, $\beta$ is the power index, and $r_0$ is the characteristic core radius.

\begin{figure}
\begin{center}
\includegraphics[width=85mm]{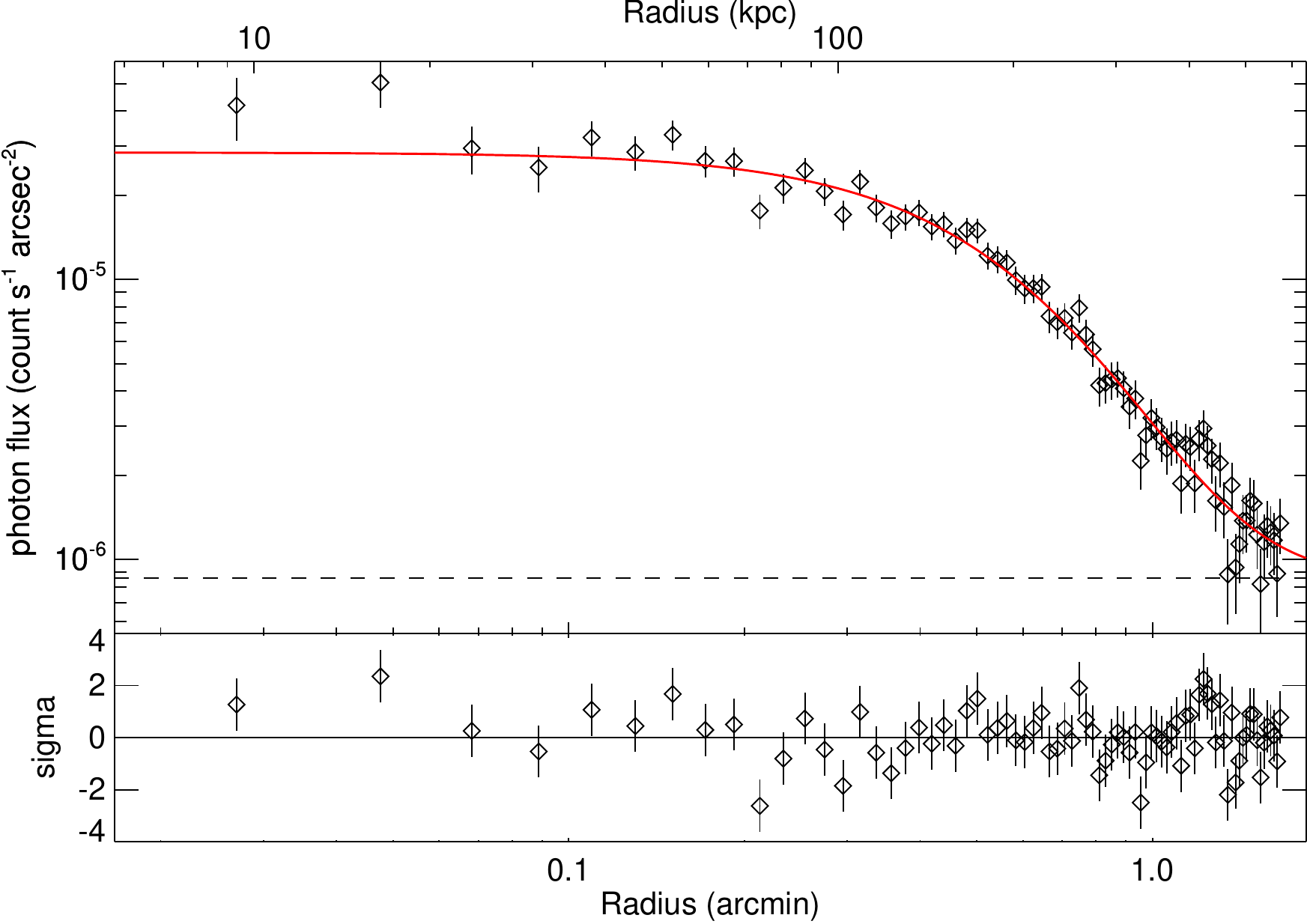}
\caption{Photon flux as a function of radius measured from the peak of the diffuse X-ray emission. The red line indicates the best-fit 1-D $\beta$-model (see Section~3.1) with an additional background term. The dashed line marks the best-fit background level. }
\label{1d_beta}
\end{center}
\end{figure}

The observed one-dimensional surface brightness profile (Fig.~\ref{1d_beta}) is computed as a function of radius by measuring the surface brightness in circular annuli centred on the peak of the X-ray surface brightness. Fitting a 1-D $\beta$-model yields $\beta=0.76^{+0.08}_{-0.10}$ and $r_0=209^{+28}_{-24}$ kpc. The goodness of fit as given by the reduced $\chi^2$  statistic is 1.06.

\begin{figure*}
\begin{center}
\includegraphics[width=165mm]{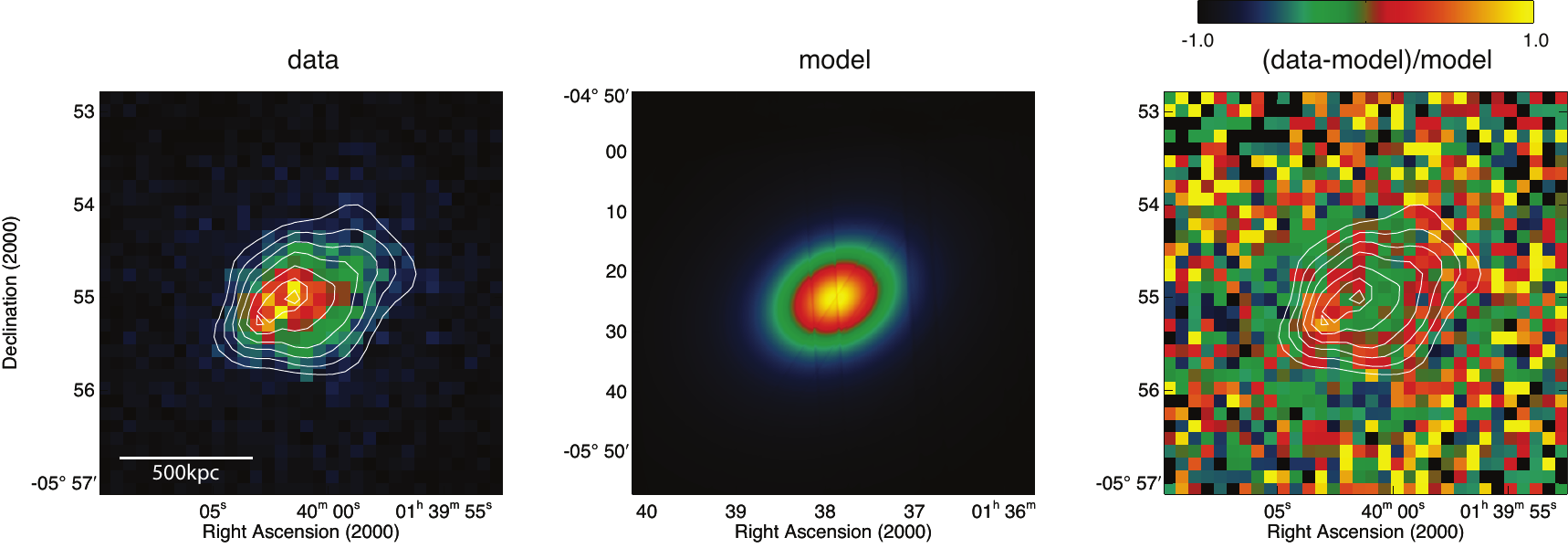}
\caption{X-ray surface brightness distribution (left), 2D $\beta$-model of the distribution (middle), and residuals (right). A constant background is included in the fit. The data and residuals are binned into 16\arcsec pixel. Contours show the adaptively smoothed \citep{ebe06} X-ray emission. Note the obvious excess in the residuals located south-east to the peak of the X-ray surface brightness. }
\label{2d_beta}
\end{center}
\end{figure*}

\begin{table}
 \caption{Best fit parameters of 2-D $\beta$-model }
  \label{table1}
   \begin{tabular}{@{}cccc}
   \hline
   $r_0$&$\epsilon\tablenotemark{1}$&$\theta$&$\beta$\\
   \hline
	203.3$^{+14.9}_{-14.0}$ kpc&$0.24\pm0.02$&$27.2^{\circ}\pm2.4$&$0.63\pm0.03$\\
   \hline
   \tablenotetext{1}{$\epsilon=1-b/a$; $a$ and $b$ are the major and minor axis, respectively.}
   \end{tabular}
   \end{table}

Since the obvious ellipticity of the X-ray emission can not be well described by the circular 1-D profile above, we use, for our two-dimensional model, a more general, elliptical $\beta$-model ({\tt beta2d} in {\it Sherpa}), which adds two more fit parameters, namely the ellipticity $\epsilon$ and the position angle $\theta$. The resulting best-fit parameters are given in Table~\ref{table1}, and the best-fit model and residuals are shown in Fig.~\ref{2d_beta}. 
Because of the poor photon statistics at large cluster-centric radii, the Cash statistic \citep{cash79} is adopted; therefore a formal goodness of fit is not available. However, a simple visual inspection of Fig.~\ref{2d_beta} proves instructive. While the overall elongation of the emission is well captured by our elliptical $\beta$-model, the deviation from concentricity in the cluster core leads to pronounced residuals when the model is subtracted from the data (right panel of Fig.~\ref{2d_beta}). Comparing the residuals to the raw data (left panel of Fig.~\ref{2d_beta}) and to the X-ray/optical overlay shown in Fig.~\ref{intro_source}, we find these residuals to be caused by the ridge-like structure in the X-ray surface brightness that spatially coincides with the core of the south-eastern (SE) sub-cluster.

\subsection{Global X-ray properties}\label{sec:global_properties}

Global X-ray properties are measured within $r_{1500}$ (0.68 Mpc or $118\arcsec$)\footnote{$r_{1500}$ is the radius at which the mass overdensity of the cluster exceeds 1500 times the critical density. Like the similarly defined radii $r_{200}$ (1.9 Mpc, 324\arcsec) and $r_{500}$ (1.2 Mpc, $205\arcsec$), $r_{1500}$ is computed as a function of k$T$ and $z$  \citep{ma08}. } which is the radius at which the signal-to-noise ratio (S/N) of the global cluster emission approximately reaches its maximum, such that extending the measurement to yet larger radii would predominantly add background (Fig.~\ref{1d_beta}). Inside $r_{1500}$, we measure a S/N of 69 for a total of 5445 net X-ray photons  in the 0.3--6\,keV band. For the spectral fit, we adopt the plasma model of \citet{mew85} combined with the photoelectric absorption model of \citet{mor83} ({\tt XSMEKAL} and {\tt XSWABS}, respectively, in {\it Sherpa}); all fits are performed simultaneously on the two observational datasets available for \source. We find $n_{\rm H}=(4.5\pm 2.0)\times 10^{20}$ cm$^{-2}$, consistent with the Galactic value of $2.85\times 10^{20}$ cm$^{-2}$ \citep{dic90};  a global temperature of $7.7^{+1.3}_{-1.0}$\,keV; and a metal abundance of $Z=0.35\pm0.12$, consistent with results obtained for other clusters at similar redshift \citep{arn92}. The goodness of fit of a single-temperature model characterized by a reduced $\chi^2$ of 1.04  confirms the reasonableness of the isothermal assumption. Adding a second temperature component to the model does not yield a better fit. In order to obtain better constraints on the ICM temperature, we freeze the absorption at the Galactic value and also set the poorly constrained global abundance to $Z=0.3$. This yields a global gas temperature of $8.5^{+0.9}_{-0.8}$\,keV.

For this final set of model parameters, the total rest-frame X-ray luminosity of the cluster within $r_{1500}$ is measured to be $1.0\times10^{45}\rm~erg~s^{-1}$, $4.9\times10^{44}\rm~erg~s^{-1}$, $9.6\times10^{44}\rm~erg~s^{-1}$, and $1.6\times10^{45}\rm~erg~s^{-1}$ (0.3--6\,keV, 0.1--2.4\,keV, 2--10\,keV and bolometric, respectively.). Extrapolating these figures to $r_{\rm 500}$ using the 2D $\beta$-model, we find values for $L_{\rm X}(<r_{\rm 500})$ of $1.3\times10^{45}\rm~erg~s^{-1}$, $6.5\times10^{44}\rm~erg~s^{-1}$, $1.3\times10^{45}\rm~erg~s^{-1}$, and $2.1\times10^{45}\rm~erg~s^{-1}$ in the same four passbands.

The normalization factor of the global spectral fit allows us to derive an estimate of the total X-ray gas mass within the region, in as much as it yields the central electron density that permits the conversion of the projected X-ray surface brightness into a physical, three-dimensional gas density. The spectral normalization of the  {\tt XSMEKAL} model is given by
$${\tt norm}={10^{-14}\over4 \pi [{D_A\over\rm{cm}}]^2(1+z)^2}\int [{n_e\over{\rm cm^3}}] [{n_H\over{\rm cm^3}}] dV.
$$

We calculate the right-hand side of this expression by numerically integrating the hydrogen number-density profile. 
The gas is assumed to be fully ionized, yielding the electron number density, $n_{\rm e}$, of 1.2 times the hydrogen number density. The hydrogen and electron density profiles are de-projected from the X-ray surface brightness fit, i.e. 2D elliptical $\beta$-model\footnote{in doing so, we assume a prolate geometry} as described in Section~\ref{sec:xsb}. Taking into account the known steepening of the X-ray surface brightness profile at large radii where our data do not constrain the mass density model, we follow \citet{vik06} and adopt a steeper slope of 0.60 at $r>0.3\times r_{200}$ and a density cutoff at $r>r_{500}$. Without these modifications, the integral over the best-fit $\beta$ model does not converge. Using this modified model, we find values for the central electron number density, $n_{\rm e}$, of $8.9\times10^{-3}\rm~cm^{-3}$ and for the total X-ray gas mass within $r_{1500}$ of $(6.55\pm0.01)\times10^{13}$\,M$_\odot$. The quoted uncertainties are the nominal $1\sigma$ errors propagated from the normalization in the spectral fitting. Note, however, that the real uncertainties are likely to be dominated by systematic errors, since we assume that (1) the gas is isothermal, (2) the gas body is a prolate spheroid with its major axis lying in the plane of the sky, and (3) the density profile is described by a modified $\beta$-model. While the assumption of isothermality on large scales is supported by our analysis of the ICM temperature distribution (Section~\ref{sec:ktmap}), the 3-D geometry of the X-ray gas is poorly constrained, and the $\beta$-model provides a poor fit to the central region of the X-ray surface brightness distribution (Figs.~\ref{1d_beta}, \ref{2d_beta}). 

\begin{figure}
\begin{center}
\includegraphics[width=80mm]{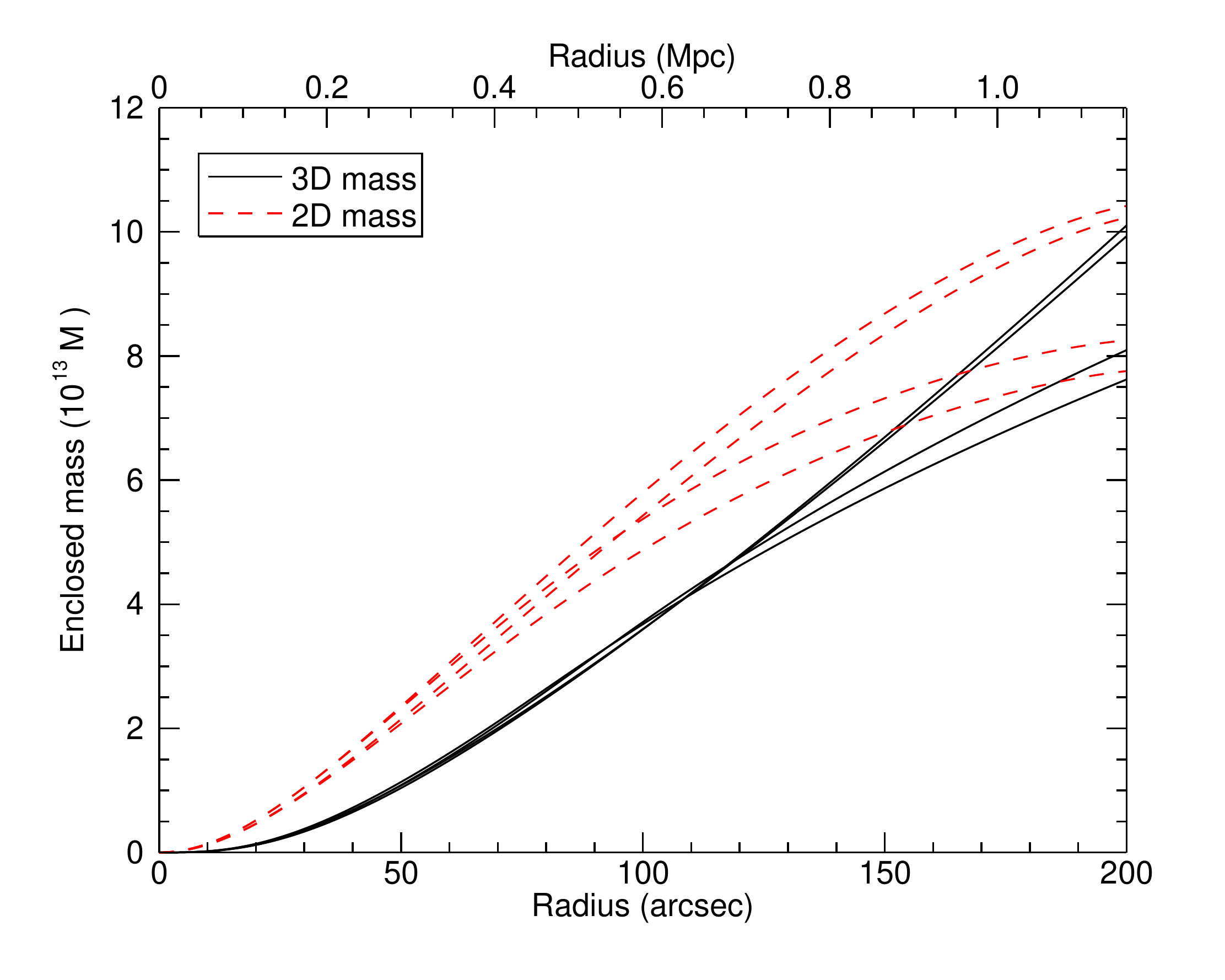}
\caption{Profiles of the cylindrically enclosed (2D) and spherically enclosed (3D) gas mass derived for different geometries and X-ray surface-brightness profile slopes. For each set of curves, the upper two assume a slope of 0.6 for the X-ray surface brightness profile at large radii, and the lower two assume 0.92; the uncertainties caused by the adopted geometry are negligible in comparison as an oblate geometry results in an only slightly higher mass than a prolate one. Details are provided in Section~\ref{sec:global_properties}.}
\label{gas_mass}
\end{center}
\end{figure}

To understand the systematic errors caused by the unknown geometry and uncertain profile at large radii, we repeat the calculation for an oblate spheroid (with a major axis that lies again in the plane of the sky) and also vary the slope of the X-ray surface brightness at  $r>0.3\times r_{200}$  from 0.60 to 0.92. The results are shown in Fig.\ref{gas_mass}. Two different types of gas mass as a function of $r$ are shown; the gas mass enclosed by an infinitely long cylinder of radius $r$ pointed toward the observer and those enclosed by a sphere of radius $r$  (the 2-dimensional and 3-dimensional gas mass, respectively). The 2D and 3D gas mass profiles illustrate that systematic effects indeed dominate at radii greater than $r_{1500}$, especially for the 2D gas mass. We find ${\rm M}( < r_{1500})\sim (5.7-6.9)\times10^{13}$\,M$_\odot$ (2D) and ${\rm M}( < r_{1500})\sim (4.5-4.7)\times10^{13}$\,M$_\odot$ (3D). At $r_{\rm 500}$ systematic errors increase dramatically, leading to a more uncertain measurement of M$_{\rm gas}(<r_{\rm 500}) = (7.8 - 10.5)\times 10^{13}$ M$_\odot$ (2D and 3D are alike due to density cutoff at $r>r_{500}$). The mass ranges quoted here are the maximum and minimum enclosed gas mass of the four different types of geometries.

The total gas mass within $r_{\rm 500}$ derived above implies a total gravitational mass for \source\ of M$_{\rm tot}(<r_{\rm 500}) =(6.8-9.1)\times 10^{14}$ M$_\odot$, assuming a baryon fraction of 0.115  \citep{man10,all08} and neglecting the mass in stars.

\subsection{X-ray temperature map}\label{sec:ktmap}

\begin{figure*}
\begin{center}
\includegraphics[width=160mm]{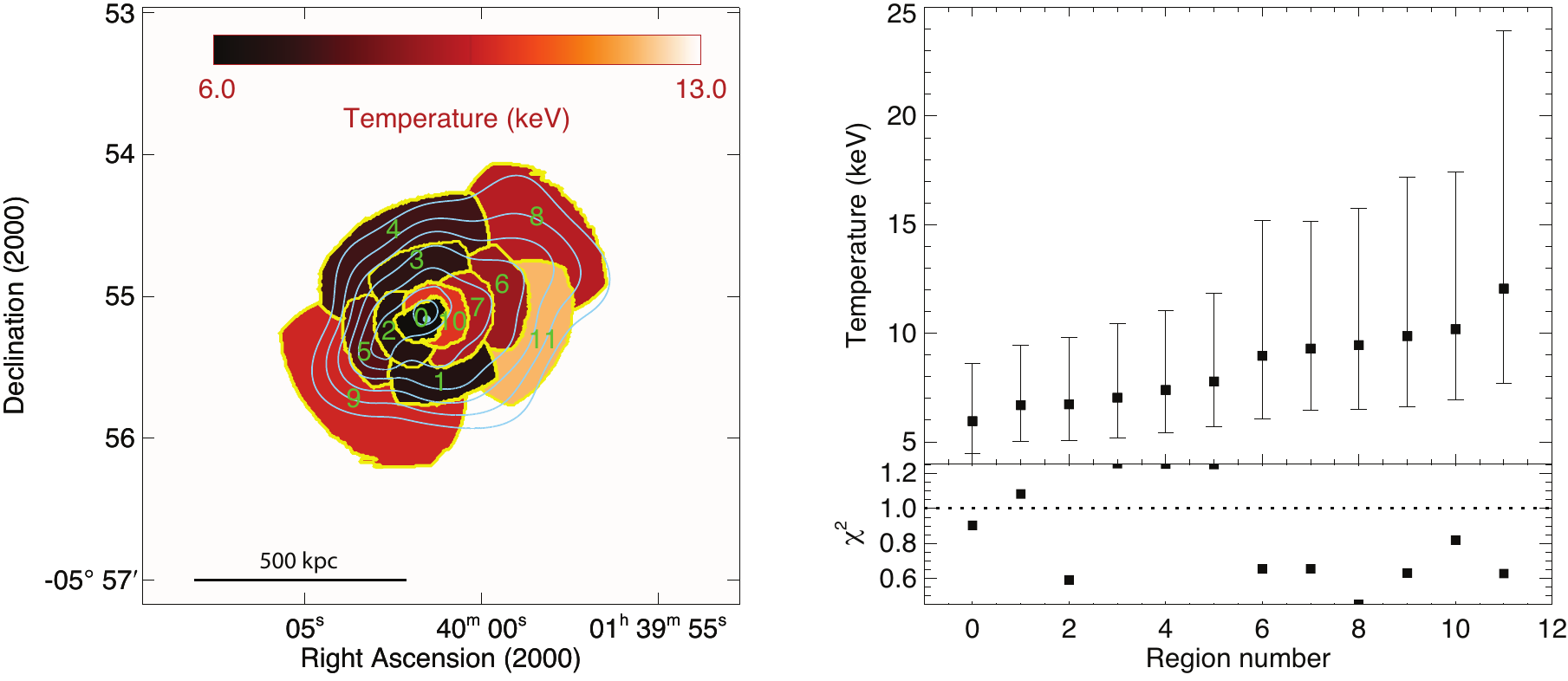}
\includegraphics[width=160mm]{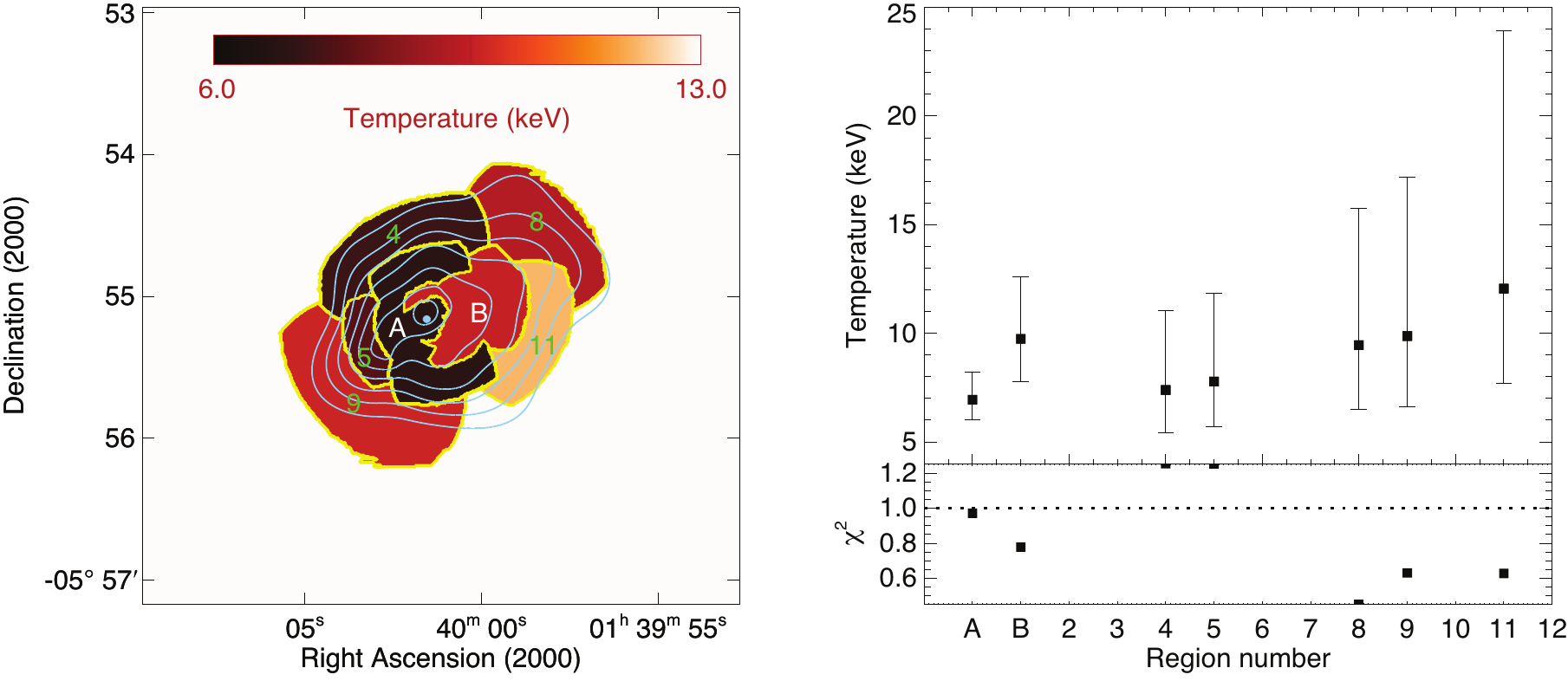}
\includegraphics[width=160mm]{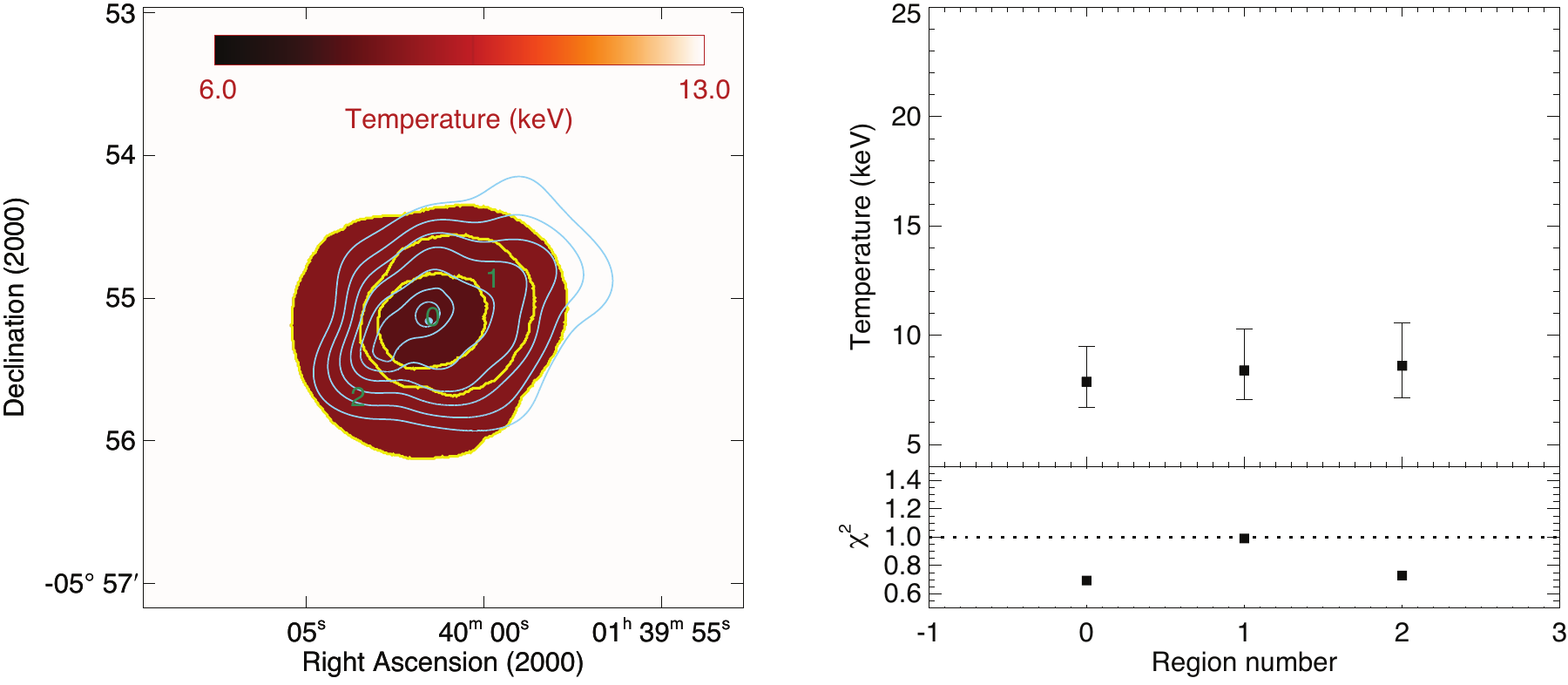}	
\caption{Left panels: temperature of the gaseous ICM estimated by fitting plasma and photoelectric absorption models to the X-ray emission in different regions (details are provided in Section~\ref{sec:ktmap}). The temperature is color coded as indicated by the color bars. Blue contours show the adaptively smoothed X-ray emission. Right panels: the best-fit temperature and reduced $\chi^2$ value for each region labeled in the left panels.  }
\label{temp_map}
\end{center}
\end{figure*}

In recognition of the limited photon statistics, we adopt the two-level binning method described in \citet{ma09} for our analysis of ICM temperature variations.
First, sub-regions are defined in the count image (0.3--6 keV) using the ``cumulative binning" algorithm \citep[{\it contbin};][]{san06}  for a given S/N threshold. Subsequently, spectral fits are performed in each region as described in Section \ref{sec:global_properties}.  Of the model parameters, only the gas temperature and the normalization are fit, while the redshift is fixed at the cluster redshift of 0.451, the absorption term is set to the Galactic value of $2.85\times10^{20}$ cm$^{-2}$ \citep{dic90}, and the metal abundance is frozen at a value of 0.3, typical of clusters \citep{arn92,bal07}. A S/N criterion of 15, corresponding to typically $\sim270$ net photons in each region, yields a first {\it high-resolution, low-fidelity temperature map}. We then manually combine adjacent regions of similar temperature, thereby creating areas comprising up to 1,300 net photons, and perform the spectral fits again to arrive at a {\it  low-resolution, high-fidelity temperature map}.

The top panels of Fig.~\ref{temp_map} show the low-fidelity temperature map and the temperatures measured in each individual region. Within the large measurement errors, the temperatures of all regions are consistent. However, careful inspection of the map suggests that the regions to the west and north-west (labeled 10, 7, 6) have consistently higher temperatures than the regions to the east and south-east (i.e., 0, 1, 2, 3). We therefore rebin these regions to increase the photon statistics. Examining the rebinned map, we find the difference in the ICM temperatures of the merged regions to be more pronounced but still statistically insignificant (i.e. regions labeled A, B).

As an alternative probe of the isothermality of the ICM we also measure the gas temperature in concentric annuli, i.e., as a function of radius. Requiring approximately 1100 net photons per annulus, we arrive at the temperature map and profile shown in the bottom panels of Fig.~\ref{temp_map}. Again we find no significant deviations from the average gas temperature of approximately 8.5\,keV, but note that variations at the level suggested by the results shown in the middle panel  of Fig.~\ref{temp_map} may still be present but undetectable with the current photon statistics of our data.

\subsection{Redshifts}

The reduced 1-D spectra of all galaxies and gravitational arc candidates targeted in our spectroscopical follow-up observations are analyzed with the IDL routine {\tt SpecPro} developed by \citet{mas11}. Spectroscopic redshifts are determined by cross-correlation against spectral templates. Given that most of the galaxies are red ellipticals, we  use the VVDS E0 template, except for five objects with strong narrow or broad emission lines for which the VVDS Starburst and SDSS Quasar templates, respectively, are adopted. The results are tabulated in Table~\ref{lris_redshift}. 

\begin{figure}
\begin{center}
\includegraphics[width=85mm]{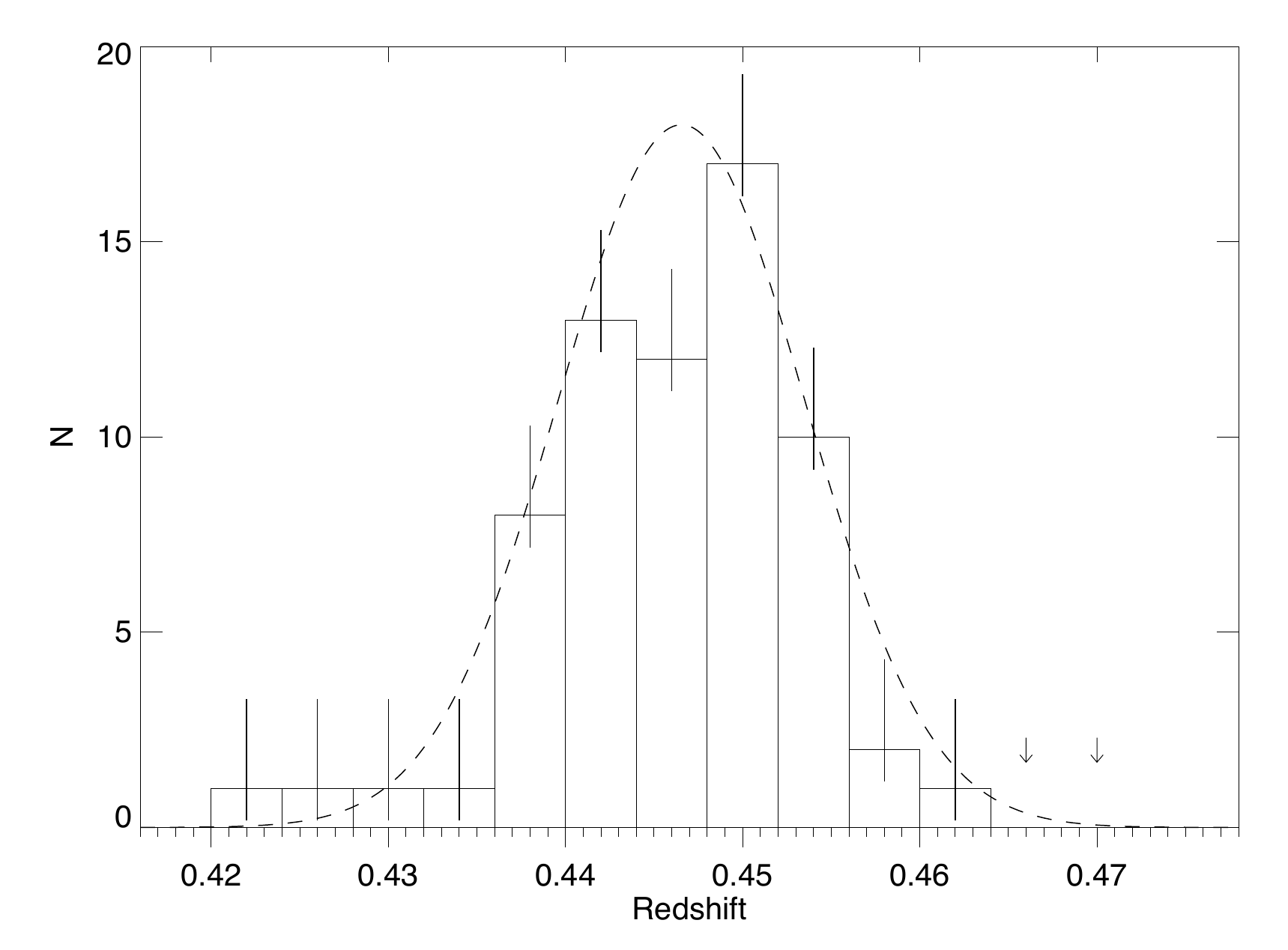}
\caption{Distribution of spectroscopic redshifts of galaxies in \source\ as determined from our LRIS and DEIMOS observations. The errors and upper limits shown assume the number of sources in each bin follows poisson statistics. The dashed line describes a Gaussian distribution with a dispersion of $875\rm~km~s^{-1}$ centred on the systemic cluster redshift of $z=0.4465$. }
\label{lris_z}
\end{center}
\end{figure}

Our redshift measurements allow us to identify 66 cluster members with redshifts between 0.423 and 0.461, two foreground objects, and four background objects, one of which is the gravitational arc (A1) discussed in more detail in Section~\ref{sec:arc}. Using the biweight and bootstrapping estimator from the ROSTAT statistics package of \citet{bee90} and converting redshifts to heliocentric radial velocities (including relativistic corrections), we calculate a mean cluster redshift of $0.4465^{+0.0011}_{-0.0009}$ and a radial velocity dispersion of $875^{+70}_{-100}\rm~km~s^{-1}$. A histogram of the redshift distribution is shown in Fig.~\ref{lris_z}. Within the limits set by the number statistics, the distribution is  well described by a Gaussian: a two-sided Kolmogorov-Smirnov test yields a probability of 88\% for the observed distribution to be drawn from a normally distributed parent distribution. Visual inspection of the distribution of redshifts on the sky (Fig.~\ref{hst}) shows no obvious correlation between the positions and the redshifts of the cluster members. 

We measure the redshifts of the two BCGs as $0.4502\pm0.0001$ for the north-western BCG (BCG1) and $0.4510\pm0.0001$ for the south-eastern BCG (BCG2). This corresponds to a difference in radial velocity of $144\pm25\rm~km~s^{-1}$.

\begin{table*}
\caption{Spectroscopic galaxy redshifts in the field of \source.}\label{lris_redshift}
\begin{tabular}{@{\hspace{2mm}}c@{\hspace{2mm}}c@{\hspace{3mm}}r@{\hspace{3mm}}c@{\hspace{2mm}}c@{\hspace{3mm}}r@{\hspace{6mm}}c@{\hspace{5mm}}c@{\hspace{2mm}}c@{\hspace{3mm}}r@{\hspace{3mm}}c@{\hspace{2mm}}c@{\hspace{3mm}}r@{\hspace{6mm}}c}
\hline
\hline
\multicolumn{3}{c}{R.A.}&\multicolumn{3}{c}{Dec.} & {\it z} & \multicolumn{3}{c}{R.A.} & \multicolumn{3}{c}{Dec.} & {\it z}\\
\hline
&&&\multicolumn{3}{c}{\bf{LRIS}}& &&&\nl
01&39&49.17&-05  &  52 & 27.78  &  0.4508  &    01 &  39 & 57.34  & -05  & 56&17.42  &  0.4383\nl
01&39&49.26&-05  &  52 &  4.20  &  0.4508  &   01 &  39 & 57.35  & -05  & 54&34.73  &  0.4368  \nl
01&39&49.88&-05  &  52 & 50.27  &  0.4436  &     01 &  39 & 57.44  & -05  & 55& 1.12  &  0.4484 \nl
01&39&50.40&-05  &  52 & 20.33  &  0.3705 &   01 &  39 & 58.03  & -05  & 54&39.57  &  0.4494 \nl
01&39&50.42&-05  &  55 & 33.95  &  0.4430 &    01 &  39 & 58.21  & -05  & 54&57.98  &  0.4610  \nl
01&39&50.70&-05  &  54 & 31.47  &  0.4563  &    01 &  39 & 58.38  & -05  & 55& 8.30  &  0.4577 \nl
01&39&52.62&-05  &  54 & 58.43  &  0.4368  &    01 &  39 & 59.10  & -05  & 56&39.69  &  0.4445 \nl
01&39&53.30&-05  &  52 & 17.88  &  0.4486  &   01 &  39 & 59.30  & -05  & 53&30.57  &  0.4455 \nl
01&39&53.97&-05  &  52 & 58.03  &  0.4471 &    01 &  39 & 59.51  & -05  & 54&58.52  &  0.4522  \nl
01&39&56.81&-05  &  54 & 24.24  &  0.4378 &    01 &  39 & 59.81  & -05  & 53&50.06  &  0.4418\nl
01&39&57.27&-05  &  54 & 46.19  &  0.4311 &    01 &  39 & 59.84  & -05  & 54&31.90  &  0.4502 \nl
01&39&57.76&-05  &  52 &  3.45  &  0.4504 &     01 &  40 &  0.13  & -05  & 56&49.62  &  0.4379 \nl
01&39&59.08&-05  &  54 & 15.97  &  0.4448 &    01 &  40 &  0.26  & -05  & 53&13.41  &  0.4430\nl
01&39&59.98&-05  &  52 & 20.03  &  0.4468 &    01 &  40 &  0.45  & -05  & 53&52.89  &  0.4556\nl
01&40& 0.22&-05  &  54 & 34.68  &  0.4481 &    01 &  40 &  0.48  & -05  & 55&46.69  &  0.4464\nl
01&40& 0.74&-05  &  55 &  1.72  &  0.4502\tablenotemark{1} &   01 &  40 &  0.83  & -05  & 54&26.79  &  0.4412 \nl
01&40& 0.83&-05  &  55 &  2.56  &  0.4504 &    01 &  40 &  1.70  & -05  & 55&19.46  &  0.4550 \nl
01&40& 1.10&-05  &  55 &  6.69  &  0.4405 &    01 &  40 &  2.11  & -05  & 54&50.22  &  0.4473 \nl
01&40& 1.47&-05  &  55 &  9.67  &  0.8729\tablenotemark{3,b} &   01 &  40 &  2.13  & -05  & 54&27.62  &  0.4451 \nl
01&40& 1.96&-05  &  55 & 13.94  &  0.4542 &   01 &  40 &  2.27  & -05  & 56&38.88  &  0.4425\nl
01&40& 2.13&-05  &  55 & 29.26  &  0.8270\tablenotemark{b} &   01 &  40 &  2.51  & -05  & 55&10.36  &  0.4354 \nl
01&40& 3.14&-05  &  55 & 20.58  &  0.4510\tablenotemark{2} &   01 &  40 &  2.75  & -05  & 53&35.74  &  0.4405 \nl
01&40& 3.84&-05  &  55 & 21.90  &  0.4547 &   01 &  40 &  2.80  & -05  & 56&48.62  &  0.4451\nl
01&40& 4.39&-05  &  56 &  6.87  &  0.4398 &    01 &  40 &  2.84  & -05  & 53&49.91  &  0.4463\nl
01&40& 5.13&-05  &  55 & 37.15  &  0.4412 &  01 &  40 &  2.86  & -05  & 55&26.48  &  0.4398 \nl
01&40& 6.02&-05  &  55 &  4.80  &  0.6096\tablenotemark{b} &   01 &  40 &  3.24  & -05  & 55& 6.47  &  0.4508\nl
01&40& 6.11&-05  &  56 & 27.71  &  0.4446 &   01 &  40 &  3.28  & -05  & 55&34.00  &  0.4504 \nl
01&40& 8.40&-05  &  54 & 58.29  &  0.4231 &     01 &  40 &  3.49  & -05  & 55&22.51  &  0.4541\nl

&&&\multicolumn{3}{c}{\bf{DEIMOS}}&&  01 &  40 &  3.89  & -05  & 55&32.22  &  0.4514\nl
    01 &  39 & 53.62  & -05  & 55& 6.64  &  0.4396 &   01 &  40 &  4.36  & -05  & 54&59.94  &  0.4508\nl
 01 &  39 & 55.39  & -05  & 55&22.26  &  0.4541&   01 &  40 &  4.96  & -05  & 56&10.91  &  0.3332\nl

  01 &  39 & 55.59  & -05  & 55&10.04  &  0.4472 &       01 &  40 &  5.11  & -05  & 54&10.70  &  0.4434\nl
   01 &  39 & 55.84  & -05  & 54&12.44  &  0.4437 &  01 &  40 &  5.37  & -05  & 56&43.50  &  0.4560\nl
  01 &  39 & 55.93  & -05  & 55& 1.28  &  0.4497 &   01 &  40 &  6.18  & -05  & 54&11.23  &  0.4418\nl
  01 &  39 & 56.90  & -05 & 54 & 35.23 & 0.9321\tablenotemark{b} &  01 &  40 &  8.51  & -05  & 54&55.35  &  0.4249\nl

    01 &  39 & 56.92  & -05  & 55&30.78  &  0.4520 & 01 &  40 &  9.21  & -05  & 54&20.65  &  0.4417\nl
   01 &  39 & 57.06  & -05  & 55&22.06  &  0.4515 \nl
   
\hline
\tablecomments{Typical errors in the measured redshifts are 0.0001 to 0.0002.}

\tablenotetext{1}{BCG1, the north-western BCG.}
\tablenotetext{2}{BCG2, the south-eastern BCG.}
\tablenotetext{3}{A1, the central gravitational arc.}
\tablenotetext{b}{Background object.}

\end{tabular}
\end{table*}

\subsection{Central gravitational arc}\label{sec:arc}

\begin{figure*}
\begin{center}
\includegraphics[width=170mm]{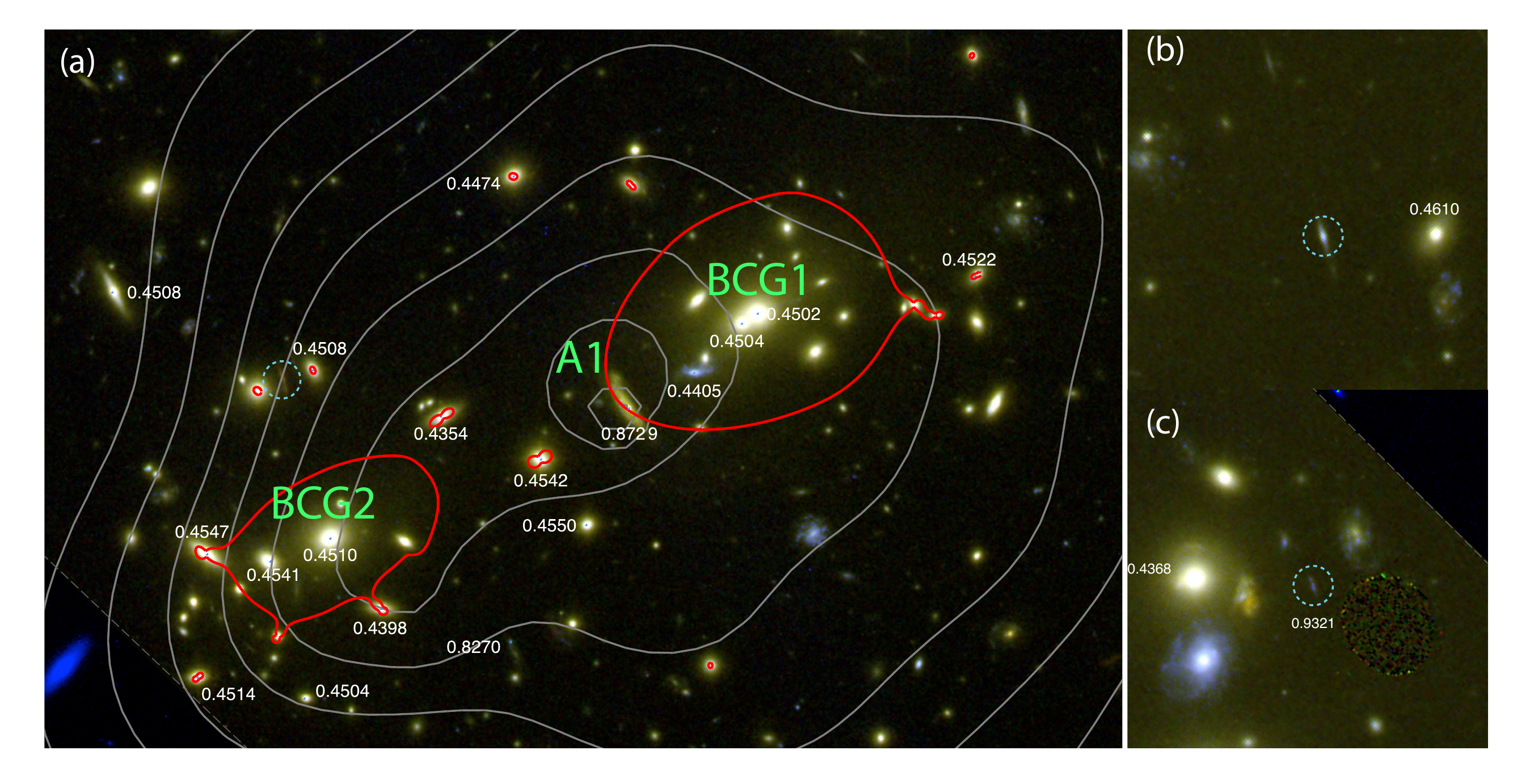}

\caption{{\it HST} colour images ({\bf(a):}$1.5\arcmin {\times} 1\arcmin$, {\bf (b) \& (c): $20\arcsec\times20\arcsec$}; blue: ACS/F606W, green: WFC3/F110W, red: WFC3/F140W). 
In {\bf(a)}, the core region of \source, grey contours show the adaptively smoothed X-ray emission and red contours show the critical lines at $z=0.87$ derived from our mass model (see Section~\ref{sec:lens_model}). The BCGs of the two sub-clusters as well as the central gravitational arc are labeled. Redshifts as measured with LRIS and DEIMOS are labeled in white in all panels. Strong lensing features compatible with our mass model are marked with blue dashed circles (one in each panel {\bf(a), (b) \& (c)}). We only successfully measured the redshift for the feature in {\bf(c)}. The lower left corner in {\bf(a)} and the upper right corner in {\bf(c)}, as marked by the grey dashed lines, were only observed with ACS. The noisy circular region lower left to the center of {\bf(c)} are dead pixels on WFC3.}
\label{hst}
\end{center}
\end{figure*}

\begin{figure*}
\begin{center}
\includegraphics[width=180mm]{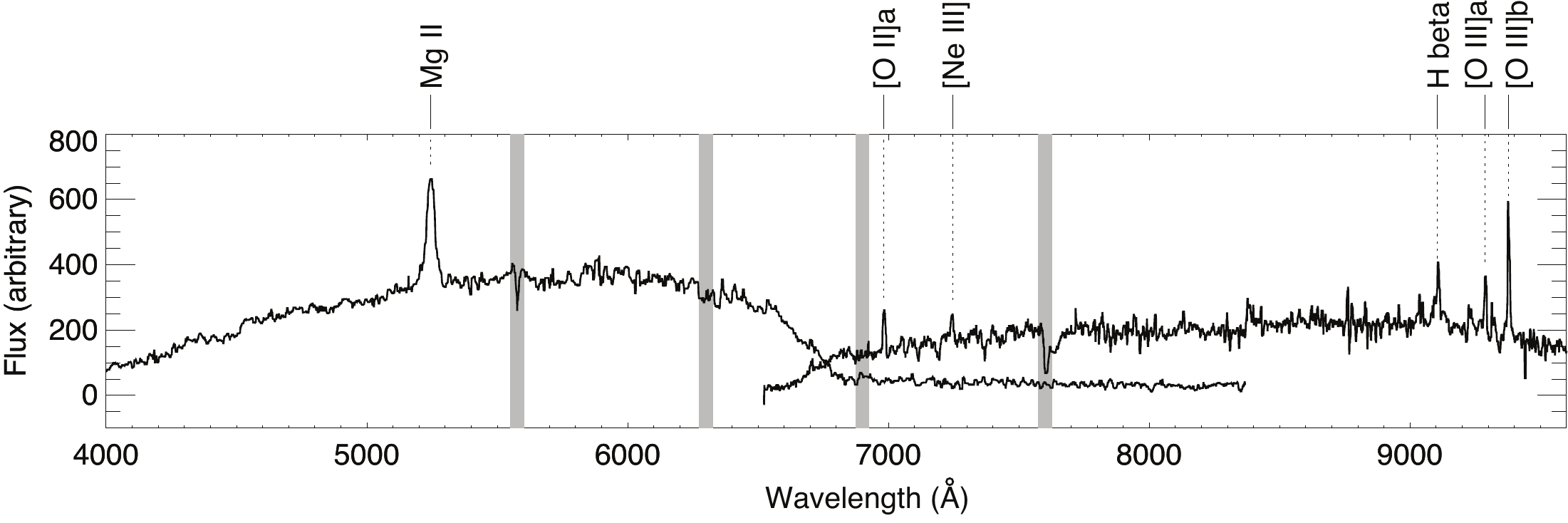}
\caption{Spectrum of A1, the gravitational arc between the two BCGs (see Fig.~\ref{hst}), as observed with LRIS. The spectra obtained with the blue and red arm of the instrument are not flux calibrated and have been median-smoothed to increase S/N. Identified emission lines are labeled.  Grey vertical bands mark the locations of absorption and emission features in the Earth's atmosphere. We derive a redshift of 0.8729 for A1, which places it well behind \source. }
\label{a1_spec}
\end{center}
\end{figure*}

The presence of a potential strong-lensing feature between the two BCGs, labeled  A1 in Table~\ref{lris_redshift}, raises the possibility of obtaining direct constraints on the mass distribution of  \source.  In the optical (Fig.~\ref{hst}), A1 appears to be a nearly straight gravitational arc at an approximately tangential orientation with respect to the two BCGs. LRIS spectra of A1 are shown in Fig.~\ref{a1_spec} with identified lines labeled. Strong emission lines in both the blue and red side of the spectrum allow an unambiguous redshift measurement of $z{=}0.8729$, placing this object well in the background of \source\ and confirming its nature as a gravitational arc. The broad line profiles and the unresolved, variable X-ray emission suggest that A1 is a Seyfert I active galactic nucleus.

In the X-ray regime, A1 coincides with a highly time-variable point source. We measure 0.3--6~keV fluxes of $(4.9^{+13.5}_{-4.9})\times10^{-16}\rm~erg~s^{-1}~cm^{-2}$  from the first {\it Chandra} observation in 2004 (too little for the source to be visible in Fig.~\ref{intro_source}), and $(1.1\pm0.3)\times10^{-14}\rm~erg~s^{-1}~cm^{-2}$  from the second one in 2010. However, the implied variability of at least a factor of 4 (but conceivably much more) over a time of several years still needs to be adjusted by the gravitational amplification by the cluster at the location of the arc. Our mass model (see below) estimates this magnification factor to be about $10\pm 2$ which reduces the variability to a modest level commonly observed in AGN \citep[e.g.,][]{markowitz03,markowitz04,pap08}.

\subsection{Mass model from strong gravitational lensing}\label{sec:lens_model}

Strong-lensing constraints are provided by the location, geometry, and orientation of A1, all of which identify it as a highly magnified tangential arc, either very close to or crossing the critical line. We attempt to reproduce the observed constraints using a series of parametric models created with the {\it Lenstool} software \citep{jul07}. 

In our model of the mass distribution, we include both smooth cluster-scale components as well as galaxy-scale components. For the latter, each cluster galaxy is assigned a small-scale isothermal elliptical halo scaled according to its luminosity (see \citealt{ric10} for more details on the modeling process). For the cluster scale mass haloes, we assume either  one pseudo-isothermal component centered on the north-western BCG, or two components centered on either BCG. Although we cannot precisely constrain the mass ratio between the main and the secondary cluster-scale components, we find that the large radius of curvature of  A1  is better reproduced by models with two haloes, with the second, south-eastern halo containing up to 50\% of the total mass of the main halo. The geometry of the critical lines (Fig.~\ref{hst}) is also compatible with other strong-lensing features in \source, found to be single images of galaxies with spectroscopic redshifts of $z{\sim}1$. 

While these simple models provide no tight constraint on the mass of the south-eastern cluster component, they yield a robust measurement of the mass of the north-western main cluster. We find  $M({<}75\,{\rm kpc})=(5.6\pm0.5) \times 10^{13}$ M$_\odot$ for the projected mass out to the radial distance marked by the location of A1. 
This value is free of systematic uncertainties as long as the mass distribution of the north-western cluster is centred on the location of BCG1 within 1\arcsec, an assumption that is supported by strong-lensing analyses of many other clusters \citep{ric10}.

\section{Discussion}

In the following we attempt to address three  questions: a) what is the inclination of the merger axis relative to the plane of the sky, and can we constrain the impact parameter of the collision, b) what is the likely total mass of \source\ and where does the system fall on the scaling relations regarding mass, X-ray temperature, and X-ray luminosity, and c) can this cluster be used to tighten the existing constraints on the self-interaction cross section of dark matter?

\subsection{Merger dynamics and geometry}
Our analysis of the imaging and spectroscopic evidence at both optical and X-ray wavelengths confirms our original hypothesis that \source\ is  an active merger of two sub-clusters. A single emission peak is observed in the X-ray regime, located between the two BCGs, and no evidence of a surviving cool core is apparent in either the X-ray surface brightness or the ICM temperature maps.  The presence of a gravitational arc with a spectroscopic redshift of $z=0.8729$ yields strong-lensing constraints on the mass distribution that favour a two-component mass model. The relative masses of the two subclusters are, however, poorly constrained but likely to fall in the range of 1:4 to 1:2. Spectroscopic redshifts  of 66 cluster members show no evidence of a bimodal distribution indicative of a line-of-sight merger and also no obvious correlation between radial velocity and position on the sky. Our tentative conclusion that the merger axis lies close to the plane of the sky will be examined more closely in the remainder of this section.

We first note that the relative radial velocity of the two BCGs  of $144\pm25\rm~km~s^{-1}$ is very small compared to the terminal velocity of BCGs in massive clusters undergoing a head-on collision. \citet{mar04} estimate the merger velocity for the Bullet Cluster to be $4500\rm~km~s^{-1}$ by calculating the free-fall velocity onto a mass concentration described by a King profile with $\rho_0\simeq2.6\times10^{25}\rm~g~cm^{-3}$ and $r_c\simeq210$ kpc. \citet{bra08}  estimate a merger velocity of ${\sim}2000\rm~ km~s^{-1}$ for MACS\,J0025.4--1222 by scaling the results of \citet{mar04} to match the velocity dispersion and  mass of their target cluster. In either case, the merger velocity is clearly supersonic\footnote{In the X-ray emitting gas of massive clusters sound travels at a speed of approximately $1300~\rm km~s^{-1}$.} owing to the deep gravitational potential well. Since the X-ray gas mass of \source\ is comparable to that of MACS\,J0025.4--1222, and the total masses of the two systems are also comparable (discussed below),  we can expect the merger velocity of \source\ to be the order of a few 1000 km s$^{-1}$. 

The observed small relative radial velocity of the two BCGs may be explained by any of the following three scenarios: (1) the two clusters are approaching each other along an axis almost aligned with our line of sight but are still so far apart that the Hubble expansion essentially cancels out the aforementioned free-fall velocity; (2) the two clusters are colliding along an axis almost aligned with the plane of the sky such that the observed difference in radial velocity represents only a small fraction of their relative velocity in three dimensions; (3) the two clusters are observed near turnaround after the primary collision and are close to at rest with respect to each other --- in this final scenario, the orientation of the merger axis is barely constrained at all. Of these three scenarios, we can firmly rule out the first one, since the required separation of several Mpc would render the system a pre-merger. Prior to first contact, however, the gaseous cores of the two clusters should align well with their BCGs which is in conflict with the observed X-ray/optical morphology shown in Fig.~1. Distinguishing between scenarios 2 and 3 is much more difficult based on the data at hand. Assuming that turnaround after the primary collision occurs approximately at the virial radius ($r_{\rm 200}$), the observed separation of the BCGs in the plane of the sky of about 250 kpc implies an inclination of the merger axis of less than 82$^\circ$ under scenario 3 --- a very weak constraint indeed. Scenario 2 (the one adopted by \citet{bra08} for MACS\,J0025.4--1222) is appealing mainly because of its simplicity; however, absent all signs of shocks in the X-ray data it is not strongly favoured by the observational evidence.

Although any conclusions regarding the impact parameter of the collision must necessarily remain somewhat speculative too, three pieces of observational evidence support the classification of \source\ as a post-collision, binary, head-on merger by \citet{man12}: (1) the excellent alignment of the X-ray emission with the line connecting the BCGs of the two subclusters; (2) the elongation of the X-ray contours along the merger axis suggested by that same line; and (3) the absence of two X-ray peaks that could be associated with the cores of the individual subclusters and the position of the single X-ray peak between the BCGs.

To summarize, we conclude that \source\ is a binary merger proceeding at close to zero impact parameter. In view of the absence of clear signs of a very recent collision (be it in the form of shocks, surviving cluster cores, or pronounced disturbance in the ICM), we propose that \source\ is observed well after the primary collision and quite possibly after turnaround. In this scenario the inclination of the merger axis with respect to the plane of the sky cannot be constrained beyond stating that it is likely to be less than approximately $80^\circ$.

\subsection{Cluster scaling relations}
The estimate of the total mass of $M(<r_{\rm 500}) = (6.8-9.1)\times 10^{14}$ M$_\odot$, combined with the global gas temperature of $8.5^{+0.9}_{-0.8}$\,keV and a 0.1--2.4\,keV luminosity of $L_{\rm X}(<r_{\rm 500})$ of $6.5\times10^{44}\rm~erg~s^{-1}$ (see Section~\ref{sec:global_properties}) places \source\ within the scattering of the Luminosity--Mass and Temperature--Mass scaling relations \citep{man10}.

\subsection{Implications for the properties of dark matter}

Although the lack of deep {\it HST} imaging prevents us from measuring the mass distribution in \source\ using weak-lensing techniques, the estimates derived from the measured gas mass and from the simple strong-lensing mass model (Sections~\ref{sec:global_properties} and \ref{sec:lens_model}) strongly suggest total masses for the two subclusters that are comparable to those of MACS\,J0025.4--1222 of $\approx2-3\times10^{14}\rm~M_\odot$ \citep{bra08}. \source\ thus has the potential to provide independent constraints on the dark matter self-interaction cross section. 

\section{Summary}

Our joint X-ray / optical analysis of \source\ based on observations with {\it Chandra} and {\it HST} finds a velocity dispersion of $\sigma=875^{+70}_{-100}\rm~km~s^{-1}$ ($n_{\rm z}=66$) as well as $L_{\rm X, bol}(<r_{\rm 500})=2.1\times10^{45}\rm~erg~s^{-1}$, k$T\sim 8.5$ keV, and $M({<}r_{\rm 500}) = (6.8-9.1)\times 10^{14}$ M$_\odot$. While the system's X-ray / optical morphology clearly identifies \source\ as a post-collision merger, there remains little observational evidence of the collision in the form of shock fronts, temperature gradients, or a bimodal redshift distribution of the cluster galaxies. We conclude that \source\ is a well advanced merger.  Although all observational evidence is consistent with the merger proceeding at minimal impact parameter and close to the plane of the sky, we are unable to formally constrain the inclination of the merger axis with respect to the plane of the sky, due to the inherent degeneracy between (radial) peculiar velocity and Hubble flow. Heuristic arguments, however, suggest that \source\ is a head-on, binary merger along an only mildly inclined axis, observed well after the primary collision and possibly after turnaround of the merger components. Our strong-lensing analysis of the system confirms the presence of two well separated mass distributions, making \source\ a potential candidate for a measurement of $\sigma/m$, the self-interaction cross section of dark matter. However, resolving the two subclusters, which are separated by only 43\arcsec (250 kpc) in projection, would pose a challenge for a weak-lensing analysis.

\section*{Acknowledgements}
This research has made use of data obtained from the Chandra Data Archive and software provided by the Chandra X-ray Center (CXC) in the application packages {\it CIAO}, {\it ChIPS}, and {\it Sherpa}.
Part of this research is based on observations made with the NASA/ESA Hubble Space Telescope, obtained from the data archive at the Space Telescope Science Institute. STScI is operated by the Association of Universities for Research in Astronomy, Inc. under NASA contract NAS 5-26555.

HE gratefully acknowledges financial support from SAO grant GO1-12153X and STScI grants GO-10491 and  GO-12166.  JR is supported by the Marie Curie Career Integration Grant 294074. 
We thank Chao-Ling Hung and Matthew Zagursky for contributions to the {\it HST} and Keck data reduction, and the UH Time Allocation Committee for their support of this project. The authors also wish to recognize and acknowledge the very significant cultural role and reverence that the summit of Mauna Kea has always had within the indigenous Hawaiian community.  We are most fortunate to have the opportunity to conduct observations from this mountain.

\end{document}